\newcommand{\ale}{\ \raisebox{-.3ex}{$\stackrel{<}{\scriptstyle \sim}$}\ }

\documentstyle{mn}

\input psfig

\title[Accretion disc boundary layers]
	{Magnetic activity in accretion disc boundary layers}
	
\author[P.J. Armitage]{Philip J. Armitage \\
	School of Physics and Astronomy, University of St Andrews, 
	North Haugh, St Andrews, Fife KY16 9SS, UK}	
 	
\begin{document}

\maketitle

\begin{abstract}
We use three dimensional magnetohydrodynamic simulations to study 
the structure of the boundary layer between an accretion disc and a 
non-rotating, unmagnetized star. Under the assumption that 
cooling is efficient, we obtain a narrow but highly variable transition 
region in which the radial velocity is only a small 
fraction of the sound speed.
A large fraction of the energy dissipation occurs in high density 
gas adjacent to the hydrostatic stellar envelope, and may 
therefore be reprocessed and largely hidden from view of the observer.
As suggested by Pringle (1989), the magnetic field energy in the boundary 
layer is strongly amplified by shear, and exceeds that in the disc by an order 
of magnitude. These fields may play a role in generating the magnetic 
activity, X-ray emission, and outflows in disc systems where the 
accretion rate is high enough to overwhelm the stellar magnetosphere. 
\end{abstract}

\begin{keywords}	
	accretion, accretion discs --- magnetic fields --- stars: winds, outflows ---
	stars: pre-main-sequence --- novae, cataclysmic variables --- MHD
\end{keywords}

\section{Introduction}

The boundary layer between an accretion disc and a slowly 
rotating star can emit up to one half of the total accretion 
luminosity (Lynden-Bell \& Pringle 1974), 
and has long been implicated as a probable site of 
variability (Pringle 1977; Papaloizou \& Stanley 1986; Kley \& 
Papaloizou 1997; Bruch 2000; Kenyon et al. 2000).
Accretion via a boundary layer is likely both in 
weakly magnetic cataclysmic variables and neutron stars, 
and in young protostars where the accretion rate is high enough to 
overwhelm the stellar magnetic field. 

Two theoretical difficulties hamper study of boundary 
layer structure. First, the dissipation 
of large amounts of energy into a narrow annulus  
requires a two-dimensional treatment of the 
thermal structure and radiation physics. Recent calculations  
of the structure of boundary layers in neutron star 
(Popham \& Sunyaev 2001) and protostellar (Kley \& Lin 1996, 1999) 
accretion have accomplished this goal.
Second, the shear and radial force balance in the boundary 
layer are qualitatively different to the Keplerian disc, 
making approximate treatments of the angular momentum 
transport (or viscosity) especially 
suspect. This aspect of the problem has been less extensively 
investigated, although it is known that the Shakura-Sunyaev 
$\alpha$ viscosity prescription (Shakura \& Sunyaev 1973) needs
to be modified to yield physically reasonable boundary 
layer solutions (Pringle 1977; Papaloizou \& Stanley 1986; 
Popham \& Narayan 1992).

This paper presents initial results from 
boundary layer calculations that dispense with approximate 
viscosity prescriptions. Instead, numerical simulations 
are used to directly resolve the physical processes that lead  
to angular momentum transport. In sufficiently well-ionized discs, 
which include the inner regions of protoplanetary discs (Gammie 1996), 
the most important process is probably turbulence driven 
by magnetorotational instabilities (Balbus \& Hawley 1991). The 
nonlinear study of these instabilities requires three 
dimensional magnetohydrodynamic (MHD) simulations, which we 
use to model the boundary layer between a geometrically thin 
accretion disc and a non-rotating, unmagnetized star. We confirm 
some aspects of prior theoretical work, but also find evidence for 
two novel effects -- magnetic activity from fields amplified in 
the boundary layer (Pringle 1989), and dissipation which is 
concentrated in relatively high density regions (Clarke \& Edwards 
1989).

\section{Numerical methods}

The ZEUS code (Stone \& Norman 1992a, 1992b, Norman 2000) is used 
to solve the equations of ideal MHD. ZEUS is an explicit, Eulerian 
MHD code, which uses an artificial viscosity to capture shocks. With 
the simplifying assumption that the fluid is isothermal (physically, 
this amounts to assuming that cooling occurs more rapidly than 
the other time-scales in the problem), the equations 
to be solved are,
\begin{eqnarray}
 { {\partial \rho} \over {\partial t} } + \nabla \cdot ( \rho {\bf v} ) 
 & = & 0 \\
 { {\partial {\bf S} } \over {\partial t} } + \nabla \cdot ( {\bf S v} ) 
 & = & - \nabla P - \rho \nabla \Phi + {\bf J} \times {\bf B} \\
 { {\partial {\bf B} } \over {\partial t} } & = & \nabla \times ( 
 {\bf v} \times {\bf B} ) \\
 P & = & \rho c_s^2,
\end{eqnarray} 
where ${\bf S} = \rho {\bf v}$, $\Phi$ is the gravitational potential, 
and the remaining symbols have their usual meanings.

The boundary layer itself is narrow. However, a larger domain is 
needed to model magnetorotational instabilities in the disc, 
which determine the structure of magnetic fields advected into 
the boundary layer region. We simulate a wedge of disc in 
cylindrical co-ordinates $(z,r,\phi$), using uniform gridding 
in both the vertical and azimuthal directions. For the radial 
direction, a non-uniform grid is employed in which the radii 
of successive grid cells are related by,
\begin{equation}
 r_{j+1} = (1 + \delta) r_j,
\end{equation}
with $\delta$ a constant. This concentrates resolution in the 
inner region of the flow. 

High resolution, especially in the radial direction, is essential. 
To simplify the problem further, we ignore the vertical component 
of gravity and consider a vertically unstratified disc. This is a 
fair first approximation to the dynamics of the flow near the disc 
midplane, though if very strong fields were to develop in the 
simulation we would have to worry that they were overestimated 
due to the neglect of buoyancy. In practice, such strong fields 
were not obtained.

The initial conditions for the simulation comprise a static, 
non-rotating and unmagnetized atmosphere, surrounded by a  
Keplerian disc with an approximately gaussian density profile.
To ensure that the initial conditions represent an accurate 
numerical equilibrium, the disc plus atmosphere structure was 
evolved in one dimension for a long period until all transients 
had died out. The resulting density and velocity profile (differing 
slightly from the input model) was then transferred to three 
dimensions, and a seed magnetic field added to the disc 
to initiate magnetorotational instabilities. Local simulations 
suggest that provided the seed field is weak, the properties 
of the final turbulent state are not strongly dependent upon 
the initial field geometry (Hawley, Gammie \& Balbus 1995, 1996). 
We use a vertical seed field in order to achieve the most rapid 
transition to turbulence, and adopt an initial ratio of the 
thermal to magnetic energy density of $\beta = 5000$.

The boundary conditions are periodic in $z$ and $\phi$, reflecting 
at $r = r_{\rm in}$ ($v_r = B_r = 0$), and set to outflow at 
$r = r_{\rm out}$. Outflow boundary conditions are achieved by 
setting all fluid variables in the boundary zones equal to 
those in the outermost active zone, together with the 
constraint that $v_r \ge 0$. 

The sound speed is set such that the ratio of the sound speed to 
the Keplerian velocity at the radial location of the boundary layer 
is $\simeq 0.1$ (precisely, $c_s = 0.1$, while $v_K = 1$ at the 
inner edge of the grid). This means that radial pressure 
gradients are small compared to the gravitational force in 
the accretion disc outside the boundary layer.
In a stratified disc, the low sound speed would 
correspond to a geometrically thin flow in which the relative 
scale height $h/r \approx c_s / v_K \ll 1$. 

\begin{table}
\begin{tabular}{lcccccc}
  Resolution & r & z & $\Delta \phi$ & $n_r$ & $n_z$ & $n_\phi$ \\ \hline
  Low & $1 \leq r \leq 5$ & $\pm 0.2$ & $45^\circ$ & 180 & 36 & 48 \\
  Medium & $1 \leq r \leq 5$ & $\pm 0.2$ & $45^\circ$ & 300 & 60 & 80 \\  
  High & $1 \leq r \leq 5$ & $\pm 0.2$ & $45^\circ$ & 480 & 60 & 90 \\      
\hline
\label{runs_table} 
\end{tabular}
\caption{Summary of the computational domain and resolution of the boundary layer runs.}
\end{table}   

Table 1 lists the parameters of the three runs discussed in 
this paper, which vary only in the numerical resolution. The 
highest resolution run has $n_r = 480$, which corresponds to 
$r_{j+1} / r_j = 1.003$. The code time units are such that 
the period of a Keplerian orbit at the inner 
edge of the grid at $r=1$ is $P = 2 \pi$.

\section{Results}

\begin{figure}
\psfig{figure=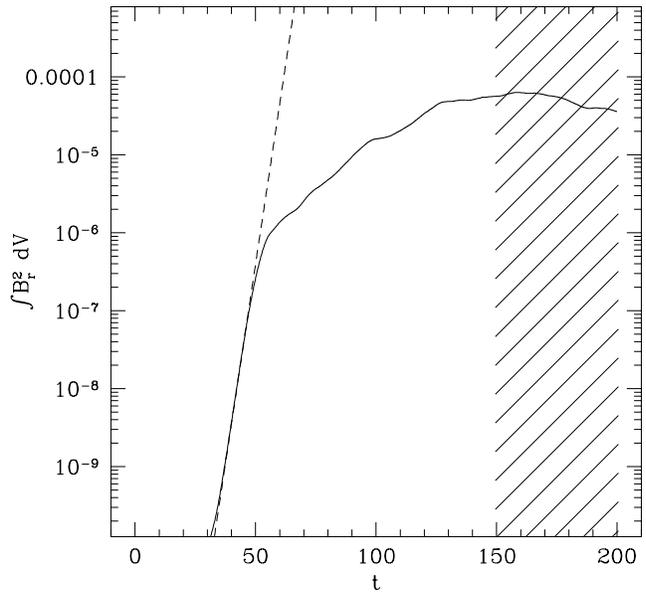,width=3.5truein,height=3.5truein}
\caption{Growth of the magnetic energy density in the radial 
         field during the high resolution simulation, integrated over the 
	 computational volume. The units on the $y$ axis 
	 are arbitrary.}
\label{f1}
\end{figure}

Radial magnetic field is generated from the initial vertical 
field by magnetic instabilities. Figure 1 shows the evolution 
of the energy density of this field component in the high resolution simulation, 
averaged over the simulation volume. An initial phase of rapid exponential 
growth is followed by a slower increase as instabilities 
in the inner disc saturate. Similar results are obtained 
in all three simulations, although saturation is reached 
marginally earlier in the higher resolution runs.
Only data from the shaded region 
near the end of the simulation, when the magnetic 
fields in the disc have reached an approximate 
steady state, is used for analysis of the boundary 
layer structure.

\begin{figure}
\psfig{figure=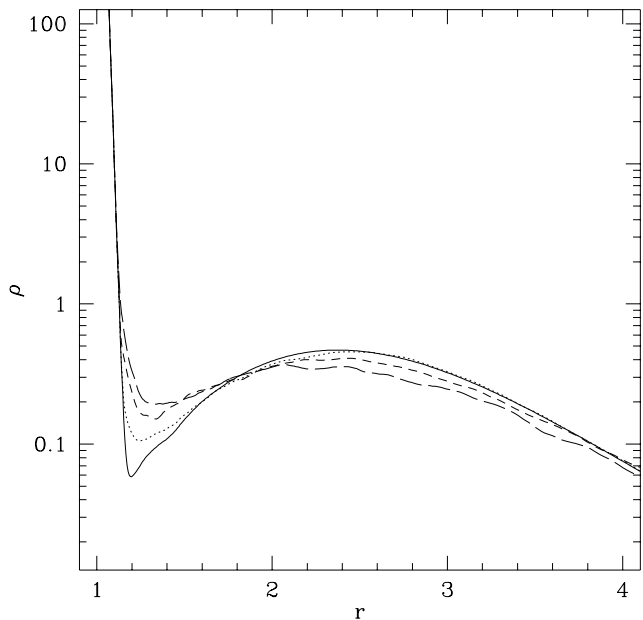,width=3.5truein,height=3.5truein}
\caption{Evolution of the disc density with time, averaged 
	over $z$ and $\phi$. Results are shown from the 
	medium resolution run at $t=50$ (solid line), 
	$t=100$ (dotted line), $t=150$ (short dashes) 
	and $t=200$ (long dashes). The initial ($t=0$) profile 
	is identical to that shown at $t=50$. Note that the 
	density at the inner edge of the grid is $\gg 10^2$ -- the 
	limits for this figure have been chosen to emphasize changes 
	in the density near the boundary layer.}
\label{f2}
\end{figure}

The magnetic fields, generated in the disc by the action 
of magnetic instabilities, lead to angular momentum 
transport. This results in a redistribution of the 
disc surface density, shown in Figure 2. As expected, 
the inner high density hydrostatic envelope remains 
static, while the centre of mass of gas in the 
disc moves inwards. This is qualitatively in 
agreement with the diffusive evolution of a 
viscous accretion disc (e.g. Pringle 1981).
Over the time-scale of the simulation, there is 
a small but clear change in the surface density 
in response to angular momentum transport in 
the disc.

\subsection{Boundary layer structure} 

\begin{figure*}
\vspace{-2.0truein}
\psfig{figure=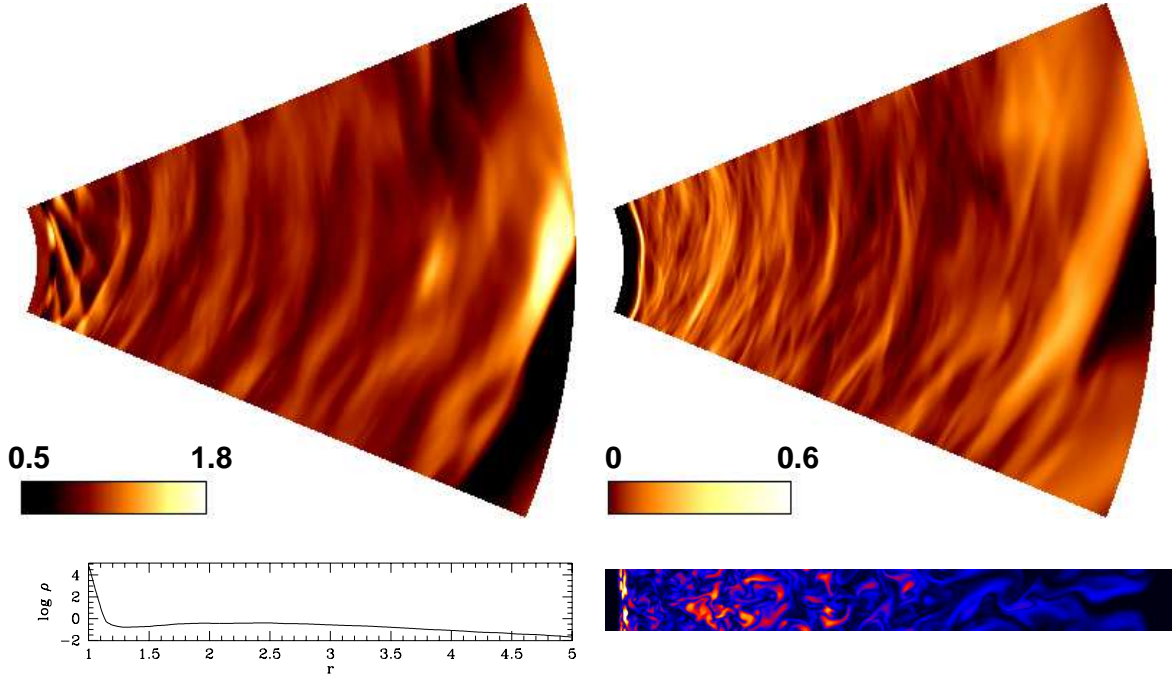,width=7.0truein,height=9.9truein}
\vspace{-4.4truein}
\caption{Results of the highest resolution run at $t=162.5$. Upper left panel: 
	map of surface density fluctuations $\Sigma(r,\phi) / \Sigma(r)$. Typical 
	fluctuations in $\Sigma$ are at the 10\% level. 
        Lower left panel: the mean density profile over the same radial range and 
	at the same time. Upper right panel: the ratio of magnetic to thermal 
	energy in the disc, averaged over $z$ in an 
	$(r,\phi)$ map. Lower right panel: energy in magnetic fields 
	({\em not} normalized to the thermal energy) in a single $(r,z)$ slice. 
	The strongest magnetic fields are typically found in the 
	boundary layer. Averaged over $z$, they reach a peak of 
	around 60\% of the thermal energy at this timeslice.}
\label{f3}
\end{figure*}

Figure 3 shows images of the magnetic fields and surface density 
fluctuations in the simulation, at a time when the inner parts 
of the disc (roughly $r \ale 4$) are fully turbulent. The magnetic 
field in the disc is predominantly toroidal, with the strongest 
fields occupying a modest fraction of the simulation volume. The 
Keplerian angular velocity in the disc means that all structures 
are strongly sheared in azimuth. In this Figure, the 
boundary layer itself is visible as a narrow stripe in the 
magnetic field map, while interior to the boundary layer the 
gas remains in its initial quiescent, almost unmagnetized state.

\begin{figure}
\psfig{figure=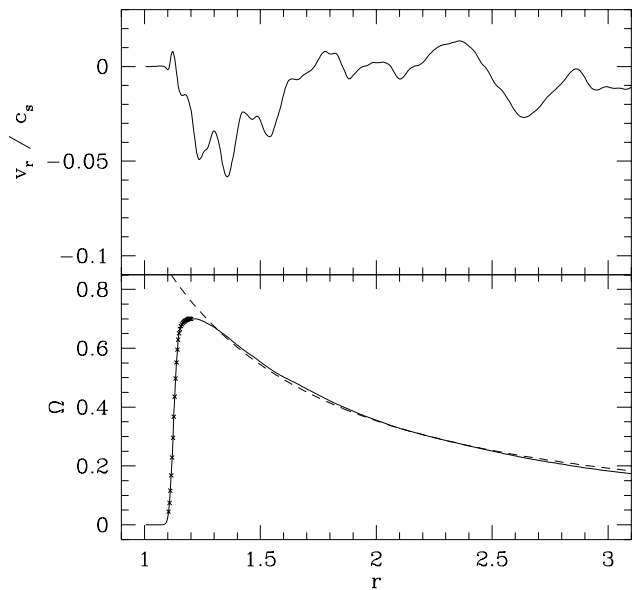,width=3.5truein,height=3.5truein}
\caption{Radial velocity $v_r$ and angular velocity $\Omega$ 
	from the high resolution simulation, averaged over 
	5 timeslices between $t=150$ and $t=200$. The dashed 
	curve in the lower panel shows a Keplerian profile 
	for the angular velocity. 
	Locations of mesh points have been plotted on the 
	angular velocity curve to indicate the resolution 
        achieved in the boundary layer region.}
\label{f4}
\end{figure}

Figure 4 shows the mean radial and angular velocity, as 
a function of radius, in the disc and boundary layer. There 
are large temporal fluctuations, especially in $v_r$, so 
the plotted quantities are averaged over $z$, $\phi$, and 
over several approximately independent timeslices from near 
the end of the simulation. We plot results from the high 
resolution run, though for the radial velocity, angular 
velocity, and surface density, consistent results are 
obtained from all three simulations.

The structure of the boundary layer seen in the 
simulations agrees with expectations based on previous 
theoretical calculations. The angular velocity, which in 
the disc is very closely equal to the Keplerian value, 
makes a smooth transition over a narrow radial region to 
the stellar value. In the highest resolution run depicted, 
this boundary layer region is resolved across about 20 
radial grid cells. The radial velocity exhibits larger 
fluctuations (even after averaging), and is actually 
largest just {\em outside} the 
boundary layer. The radial velocity remains highly 
subsonic, $v_r \ale 0.05 c_s$, at all radii, as predicted 
using arguments based on causality (Pringle 1977; 
Popham \& Narayan 1992).

\subsection{Magnetic fields in the boundary layer}

\begin{figure}
\psfig{figure=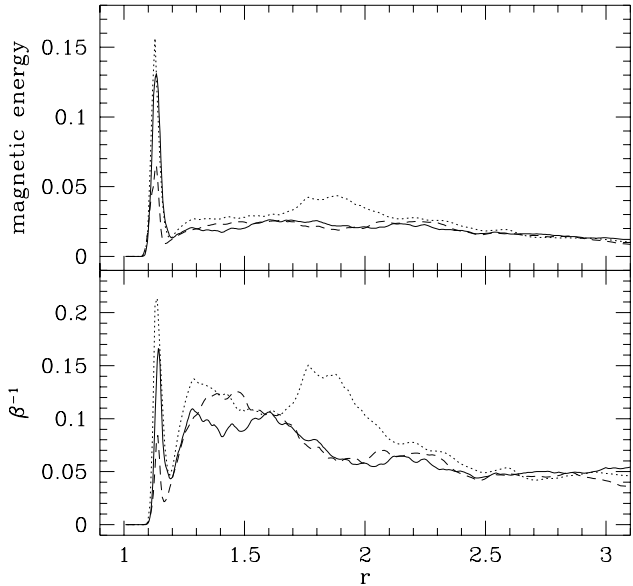,width=3.5truein,height=3.5truein}
\caption{Strength of the magnetic fields in the flow. The 
	upper panel shows the magnetic energy density as a 
	function of radius in arbitrary units, the lower 
	panel the ratio of the magnetic energy to the 
	thermal energy. Results from the low (dotted line), 
	medium (dashed line) and high resolution (solid 
	line) runs are plotted, in each case averaged 
	over several timeslices. The spike in magnetic 
	energy at $r \simeq 1.15$ corresponds to the 
	location of the boundary layer.}
\label{f5}
\end{figure}

Figure 5 shows the magnetic energy density in the 
simulations, both in absolute terms and as a 
fraction of the thermal energy of the gas. In 
absolute terms, the strongest fields by far are 
obtained in the boundary layer. The magnetic energy 
density there exceeds that in the disc at larger radii 
by roughly an order of magnitude. This is an explicit 
demonstration of the amplification of magnetic fields 
by the strong shear in the boundary layer  
(Pringle 1989). It is also analagous to the situation 
inside the last stable orbit around black holes, where 
it has similarly been suggested that the presence of strong 
shear can amplify magnetic field energy relative 
to other energies in the system (Krolik 1999).

As a fraction of the thermal energy, there is a smaller 
but still pronounced spike in magnetic energy at the 
location of the boundary layer. We obtain magnetic field 
energies in the boundary layer that are between 10\% and 
20\% of the thermal energy, compared to values in the disc 
of $\sim 5\%$ well away from the boundary layer region. Some 
individual timeslices -- for example the one shown in 
Figure~3 -- yield boundary layer fields that are 
substantially stronger still.

Unlike in the case of the radial and angular velocity, 
it is clear from Figure 5 that the magnetic field energy 
in the boundary layer has {\em not} converged at the 
highest resolution attained in the simulations (the 
magnetic fields in the disc, on the other hand, are 
consistent in the medium and high resolution runs). 
There is no clear trend with resolution, but the 
highest resolution run generates boundary layer 
fields that are substantially stronger than 
those in the medium resolution run. Since 
numerical reconnection at the grid scale is 
bound to artificially destroy magnetic fields, 
the conservative view is that 
the simulations demonstrate only that the 
boundary layer fields will be substantially stronger 
than those in the disc. Still stronger boundary layer 
fields, perhaps approaching or exceeding equipartition 
with the thermal energy (Pringle 1989), remain a 
possibility.

\subsection{Dissipation}

\begin{figure}
\psfig{figure=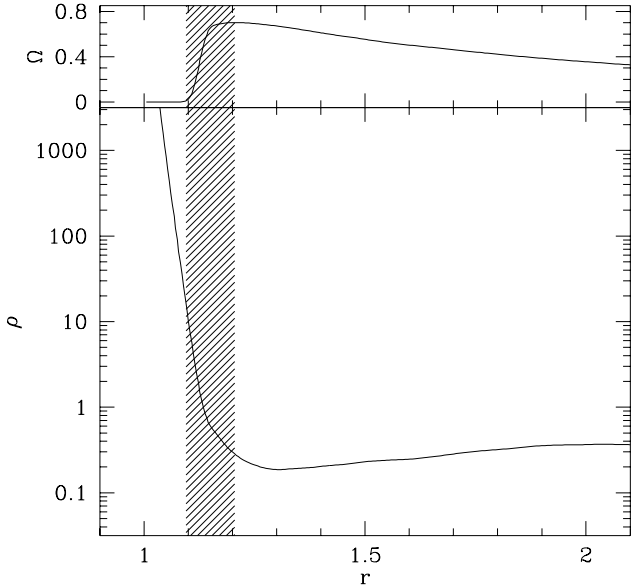,width=3.5truein,height=3.5truein}
\caption{Location of the boundary layer. The region of the 
	flow with ${\rm d} \Omega / {\rm d} r > 0$ and 
	$\Omega > \Omega_*$ (shaded 
	region) corresponds to relatively high density gas 
	adjacent to the outer part of the static atmosphere.}
\label{f6}
\end{figure}

There is no energy equation, and hence no explicit dissipation, 
in these isothermal simulations. In a more realistic 
description, however, there would be a large accretion 
luminosity arising from dissipation of the rotational 
energy of the gas in or near the boundary layer region. 
It is therefore interesting to note that in the 
simulations, the boundary layer (defined here as 
the region where ${\rm d} \Omega / {\rm d} r > 0$, 
and shown as the shaded band in Figure 6) lies in 
relatively high density gas that merges indistinguishably 
into the hydrostatic envelope that represents the star. 
This, of course, is only a restatement of the fact that 
angular momentum transport within the boundary layer is 
inefficient. As a result, the radial velocity in the boundary 
layer is very small -- smaller in fact than in the disc 
immediately outside the boundary layer, and the density high. 
If these results hold also for more realistic boundary layers 
(which will be radially broader due to thermal effects), then 
they imply that a significant fraction of the 
boundary layer emission could be intercepted and 
reprocessed by the accreting object. The boundary 
layer luminosity would then be, at least partially, 
hidden from view of the observer.

\section{Discussion}

For computational reasons, and for the sake of simplicity, 
the simulations presented here assume that the accreting 
star is unmagnetized. This is not generally true, so we 
summarize here the conditions and systems for which it 
is a reasonable approximation. 

The magnetic field of the accreting object, if 
it is strong enough, will disrupt the inner accretion disc 
and channel infalling matter along the field lines 
to the stellar surface (Pringle \& Rees 1972; Miller \& 
Stone 1997). Although the details are model dependent, 
the radius $r_{\rm m}$ at which the magnetic field will 
disrupt the disc is expected to be comparable to the 
spherical Alfven radius (K\"onigl 1991),
\begin{equation}
 { r_{\rm m} \over r_* } = \beta_{\rm t} 
 \left( { {B_*^4 r_*^5} \over {G M_* \dot{M}^2} } \right)^{1/7}.
\end{equation}
Here $B_*$ is the surface magnetic field strength (assumed 
dipolar), and $\beta_{\rm t}$ is a scaling factor of the order of
unity whose value depends upon the adopted model for the 
star-disc interaction (e.g. Ghosh \& Lamb 1978; Shu et al. 1994).

Setting $r_{\rm m} / r_* = 1$, and adopting magnetic field 
strengths and stellar radii appropriate for T Tauri stars
(Guenther et al. 1999), the minimum accretion rate required to 
disrupt the disc is,
\begin{eqnarray}
 \dot{M}_{\rm crit} \simeq 2.5 \times 10^{-6} 
 \left( { \beta_{\rm t} \over 0.5 } \right)^{7/2} 
 \left( { M_* \over M_\odot } \right)^{-1/2} \nonumber \\
 \times \left( { B_* \over {1 \ {\rm kG}} } \right)^2  
 \left( { r_* \over {3 r_\odot} } \right)^{5/2} \ M_\odot {\rm yr}^{-1}.
\end{eqnarray}  
For a sample of classical T Tauri stars in the Taurus molecular 
cloud, Gullbring et al. (1998) estimated mass accretion rates 
in the range $10^{-9} \ M_\odot {\rm yr}^{-1} \ale \dot{M} \ale 
10^{-7} \ M_\odot {\rm yr}^{-1}$. Accretion at these rates will 
result in the inner disc being disrupted by the stellar 
magnetosphere, rather than reaching the star and forming 
a boundary layer. Numerous observations support this 
conclusion (Najita et al. 2000).

In younger pre-main-sequence stars, however, the accretion rates are higher.
At early times, the accretion rate through the disc is expected to be 
of the order of $c_s^3 / G \sim 10^{-5} \ M_\odot {\rm yr}^{-1}$, 
where $c_s$ is here the sound speed in the collapsing cloud 
(e.g. Shu 1977; Basu 1998). Accretion rates of this order, 
along with (perhaps) weaker stellar fields, make 
boundary layer accretion more likely. Observationally, 
magnetic activity can certainly be present long before 
the optically visible Classical T Tauri stage.
X-ray emission, indicative 
of powerful magnetic activity, is observed in some (though by 
no means all) of the youngest Class 0 and Class I sources 
(Feigelson \& Montmerle 1999; Montmerle et al. 2000; Carkner, 
Kozak \& Feigelson 1998; Tsuboi et al. 2001).
Outflows, which provide less direct
evidence of the presence of magnetic fields, are often 
powerful even in Class 0 sources (Bontemps et al. 1996). 
Boundary layers could play a role in some of these 
phenomena.

Identical arguments apply to white dwarf and neutron star 
accretion. Adopting parameters appropriate to a dwarf nova 
in outburst, the dipole magnetic field required to just disrupt 
the disc is,
\begin{eqnarray}
 B_{\rm crit} \simeq 10^5 
 \left( { \beta_{\rm t} \over 0.5 } \right)^{-7/4}
 \left( { M_* \over M_\odot } \right)^{1/4}
 \left( { \dot{M} \over {10^{-8} \ M_\odot {\rm yr}^{-1} } } \right)^{1/2} \nonumber \\
 \times \left( { r_* \over {5 \times 10^8 \ {\rm cm}} } \right)^{-5/4} \ {\rm G}.
\end{eqnarray} 
This is not an enormous field, even though we have used an outburst accretion rate.  
In quiescence, typical accretion rates are much lower, and correspondingly 
weaker magnetic fields would suffice to disrupt the disc. Hence, in {\em quiescent} dwarf novae
magnetic fields may well be strong enough to disrupt the inner accretion 
disc (Livio \& Pringle 1992). Conversely, in supersoft X-ray sources 
(van den Heuvel et al. 1992), 
cataclysmic variables with higher accretion 
rates, and dwarf novae during outburst, boundary layers are likely to exist for 
relatively weakly magnetic white dwarfs. The results presented here 
suggest that emission from the boundary layer will be highly variable, 
but may be partially hidden from observational view.

\section{Conclusions}

In this paper, we have presented MHD simulations of the boundary 
layer between an accretion disc and a non-rotating, unmagnetized 
star. Although in these initial simulations drastic simplifications 
have been made in other aspects of the disc model, the inclusion of 
MHD allows us to resolve the physics underlying 
angular momentum transport in the disc and boundary layer 
region -- one of the main areas of uncertainty in 
previous models. Three main results emerge from the simulations:
\begin{enumerate}
\item[(i)]
The basic structure of the boundary layer agrees with 
that predicted previously. The boundary layer is highly 
variable, narrow, and the radial velocity subsonic.
\item[(ii)]
Magnetic fields generated in the disc are amplified 
by the shear in the boundary layer. Unless the star 
itself has a strong field, this means that the strongest magnetic 
fields in the star-disc system will be in the 
boundary layer. We derive magnetic energy densities 
that are a few tenths of the thermal energy. Resolution 
limitations mean that this is probably a lower limit to the 
true field strength.
\item[(iii)]
Angular momentum transport in the boundary layer 
is inefficient, and as a result 
dissipation occurs primarily in relatively high density 
gas. Radiation originating from the boundary layer may therefore 
be partially `buried' in the stellar envelope.
\end{enumerate}

These results, if they apply also to boundary layers 
with more realistic thermal structures, suggest that 
in systems where boundary layers are present, they 
will be an important site of 
magnetic activity. Magnetic fields in the 
boundary layer could be important for producing the 
observed X-ray emission in these systems, and more 
speculatively could play a role in the formation of 
outflows. 

\section*{Acknowledgements}

This work was begun during a visit to JILA, and I thank 
Chris Reynolds, Anita Krishnamurthi and Mitch Begelman for their 
hospitality. I also enjoyed  
many useful discussions with participants at the G\"ottingen and UKAFF1 
conferences. Computations made use of the UK Astrophysical 
Fluids Facility.

\end{document}